# Ambipolar Light-Emitting Transistors on Chemical Vapor Deposited Monolayer MoS$_2$

*Evgeniy Ponomarev, Ignacio Gutiérrez-Lezama, Nicolas Ubrig, and Alberto F. Morpurgo\**

DQMP and GAP, Université de Genéve, 24 quai Ernest Ansermet, CH-1211 Geneva, Switzerland

**KEYWORDS.** MoS$_2$, light-emitting transistor, ionic liquid gating, ambipolar transport, CVD

**ABSTRACT.**

We realize and investigate ionic liquid gated field-effect transistors (FETs) on large-area MoS$_2$ monolayers grown by chemical vapor deposition (CVD). Under electron accumulation, the performance of these devices is comparable to that of FETs based on exfoliated flakes. FETs on CVD-grown material, however, exhibit clear ambipolar transport, which for MoS$_2$ monolayers had not been reported previously. We exploit this property to estimate the bandgap $\Delta$ of monolayer MoS$_2$ directly from the device transfer curves, and find $\Delta \sim$ 2.4-2.7 eV. In the ambipolar injection regime, we observe electroluminescence due to exciton recombination in MoS$_2$, originating from the region close to the hole-injecting contact. Both the observed transport properties and the behavior of the electroluminescence can be consistently understood as due to the presence of defect states at an energy of 250-300 meV above the top of the valence band, acting as deep traps for holes. Our results are of technological relevance, as they show that devices with useful opto-electronic functionality can be realized on large-area MoS$_2$ monolayers produced by controllable and scalable techniques.

**TEXT.**

Monolayers of semiconducting transition metal dichalcogenides (TMDs) based on molybdenum and tungsten are at the focus of considerable attention because their electronic properties enable the investigation of interesting physical phenomena and offer a significant potential for future applications[1,2]. These monolayers are direct band-gap semiconductors that interact strongly with light, properties that make them attractive for the realization of opto-electronic devices[3]. Indeed, a variety of structures – such as photodetectors[4], solar cells[5,6], light-emitting diodes[5–8] and light-emitting transistors[9] – has been investigated and shown to be promising either because of their particularly good performance (e.g., very high sensitivity of photo-detectors[4]) or of their rather unique functionality (e.g., the possibility to control the emission of circularly polarized light[10]).

So far, the vast majority of opto-electronic devices based on semiconducting monolayer TMDs has relied on individual flakes exfoliated from bulk crystals. Nevertheless, the development of a viable technology will require the realization of structures produced through a controllable and scalable process. Progress in this direction is well on its way, as the growth of large-area, high-quality atomically thin layers (down to monolayers) of MoS$_2$[11–18], MoSe$_2$[19,20], WS$_2$[21] and WSe$_2$[22] has already been demonstrated by means of chemical vapor deposition (CVD). CVD-grown atomically thin layers of semiconducting TMDs have indeed started to be used for the realization of integrated circuits[23], gas- and photosensors[24], and flexible transistors[25]. Only limited work, however, has focused on opto-electronic applications, possibly because most opto-electronic functionalities rely on the occurrence of ambipolar transport, which is extremely difficult to observe experimentally in CVD-grown monolayers of semiconducting TMDs. Indeed, ambipolar transport in CVD-grown monolayers has been reported for the first time only recently in $WSe_2$,[26] and its observation required sweeping the gate voltage by approximately 200 V, a range by far not compatible with practical applications.

Here, we report on the use of high-quality, large area CVD-grown $MoS_2$ monolayers to realize ionic liquid (IL) gated field effect transistors (FETs) that work both in electron and hole accumulation. We exploit these ambipolar FETs to perform two different types of experiments. First, we take advantage of the very large capacitance of the ionic liquid gate to quantitatively estimate the single particle band gap $\Delta$ of $MoS_2$ monolayers directly from the FET transfer curve, a measurement that could not be done until now because – contrary to the case of thicker $MoS_2$ layers[27] – ambipolar transport in monolayer $MoS_2$ had not been reported yet. From measurements on different devices, we obtain values of $\Delta$ ranging from 2.4 to 2.7 eV. Secondly, we drive our ionic liquid gated FETs in the ambipolar injection regime, in which charge carriers of opposite polarity are injected at the source and drain contacts. In this regime, we detect electroluminescence at the energy corresponding to recombination of excitons in monolayer $MoS_2$, from the region close to the hole injecting contact. A detailed analysis of the spectra of the emitted light and of their evolution with applied bias shows – consistently with the result of gate-dependent transport measurements – that the confinement of the light emission near the contact region is caused by impurity states broadly distributed in an energy range of 250-300 meV above the top of the valence band, acting as deep traps for holes. Our results lead to an internally consistent microscopic scenario accounting for the observed transport properties and electroluminescence of monolayer MoS$_2$. Additionally, they demonstrate the possibility to employ large-area CVD-grown MoS$_2$ monolayers to realize ambipolar devices enabling light emission and operating on a voltage scale compatible with practical applications, a conclusion that has technological relevance for the development of opto-electronic devices based on atomically thin semiconducting TMDs.

Large-area MoS$_2$ monolayers are grown by CVD on sapphire substrates through chemical reaction of a molybdenum containing precursor (MoO$_3$) and sulfur, adapting a protocol reported in Ref. 18. In short (see Ref. 18 for details), crucibles with the precursors are loaded into a tube furnace, so that the MoO$_3$ powder is positioned in the middle of the furnace and the sulfur approximately 20 cm upstream. The substrate is placed "face-down" on top of the crucible containing MoO$_3$. The growth is done under a constant flow of Ar (75 sccm) with the temperatures of MoO$_3$ and sulfur reaching 700 °C and 250 °C respectively. As shown in Ref. 18, this process – in combination with the use of polished c-plane sapphire substrates – enables the epitaxial growth of MoS$_2$ monolayer with a very small number of tilted grain boundaries. An image of a synthesized MoS$_2$ monolayer is shown in Fig.1a (the red arrows point to the edges of the film).

The monolayer nature and the uniformity of the material are assessed by optical characterization. Fig. 1b shows a Raman spectrum of a CVD-grown MoS$_2$ monolayer (red curve) together with the spectrum of a bare sapphire substrate (blue curve). The peaks at 384.4 cm$^{-1}$ and 404.6 cm$^{-1}$ (see Fig. 1c), only visible after the growth process, correspond to the $E_{2g}^1$ and the $A_{1g}$ vibrational modes characteristic of crystalline MoS$_2$. Their precise energy provides a first indication that the material is indeed a monolayer[28]. This is confirmed by the photoluminescence (PL) spectrum shown in Fig.

1d [29,30] (purple curve). A peak at approximately 660 nm (1.88 eV) is seen, due to the recombination of so-called A-excitons, which matches the peak measured on exfoliated $MoS_2$ monolayers (green curve; similarly to what reported previously[17,18], the higher energy shoulder originating from the B-exciton transition that is seen in the exfoliated monolayer is not detected in the CVD-grown material). We conclude that both Raman and PL measurements show that the material grown is monolayer $MoS_2$.

To probe the material homogeneity we map the PL signal by illuminating as-grown monolayers locally with a laser beam (the spot size is approximately 1 μm in diameter). We focus on the energy of the A-exciton peak and its full width at half maximum (FWHM), which are sensitive to defects[31] and impurities[32]. Maps of these two quantities are shown in Fig. 1e, f. Except for individual points (originating from a glitch in the measurements), the A-exciton energy ranges from 657.3 to 662.6 nm (i.e., from 1.871 to 1.887 eV) and the FWHM from 22.1 to 26.6 nm (i.e., from 63.5 to 76.4 meV). We compare these ranges with the spread measured on five distinct exfoliated $MoS_2$ monolayers, for which the peak position varies in the range 657.8-672.5 nm (1.844-1.885 eV) and the FWHM ranges from 23.6 to 36.9nm (67.5-101.1 meV). Our large-area CVD-grown monolayers, therefore, exhibit smaller spread in optical properties than the sample-to-sample variations found in individual $MoS_2$ monolayers exfoliated from bulk crystals. We conclude that the homogeneity of our CVD-grown $MoS_2$ monolayers is satisfactory.

Electronic transport measurements are done using FETs realized on as-synthesized material (see Fig. 2a for a schematic illustration of the devices and the inset of Fig. 2b inset for an optical microscope image of an actual device). The gold electrodes used to measure transport, a planar gate electrode and an additional contact acting as reference electrode are realized using electron-beam lithography, metal evaporation, and lift-off. A subsequent annealing step in an inert Ar atmosphere is performed at $T = 200$ °C to lower the contact resistance[33]. A small droplet of ionic liquid (EMI-TFSI) is deposited on top of the devices[34–36] just before loading the them into a vacuum chamber ($p \sim 10^{-6}$ mbar), where they are kept at room temperature overnight (to pump out humidity and oxygen) before starting the electrical measurements. Using FETs realized in this way allows us to measure transport without the need to mechanically transfer the $MoS_2$ monolayers onto a different substrate, a step that can easily introduce defects detrimental for the observation of ambipolar transport.

Fig. 2b-d show representative transfer curves (current $I_{SD}$ as a function of gate voltage $V_G$ for different values of source-drain bias $V_{SD}$) and output curves (current $I_{SD}$ as a function of $V_{SD}$ for different $V_G$ values) of our devices measured under electron accumulation. Virtually no hysteresis is observed upon increasing the gate voltage and sweeping it back, as long as the maximum value of $V_G$ remains within a few hundreds milliVolts from threshold (see Fig. 2b). Increasing the range over which $V_G$ is swept results in a larger hysteresis (see Fig.2c) and is accompanied – upon repeated cycling of $V_G$ – by a shift in threshold voltage to lower values, in complete analogy to what is commonly observed in IL-gated devices realized on exfoliated flakes. The value of the square resistance $R^\square$ for $V_G \sim 1$ V above threshold is approximately 10 k$\Omega$, only slightly larger than the value measured on exfoliated $MoS_2$ monolayers at a comparable gate voltage. At this same gate voltage we measure the carrier density from Hall-bar shaped devices identical to the ones discussed here (see Supplementary Information for details) and obtain $n \sim 10^{14}$ cm$^{-2}$, from which we extract the room-temperature electron mobility to be $\mu = \frac{1}{R^\square n e} \sim 10$ cm$^2$ V$^{-1}$ s$^{-1}$. This value is comparable to what has been reported earlier for exfoliated and (top-quality) CVD MoS$_2$ monolayers [15,16,37].

The current upturn seen past the saturation regime at large positive $V_{SD}$ and small (to moderate) $V_G$ values is worth commenting (see Fig. 2d for $V_{SD}$ = 1.5 V or larger). It is an experimental manifestation of the onset of the ambipolar injection regime, in which electrons and holes are simultaneously injected at the two opposite contacts in the transistor[38], and it implies the occurrence of ambipolar transport. This observation is interesting because – contrary to monolayers of other semiconducting TMDs – for MoS$_2$ monolayers the manifestation of ambipolar conduction had not been reported yet (neither on exfoliated nor on CVD-grown material). To confirm the occurrence of ambipolar conduction we measure the full transfer characteristic by sweeping $V_G$ up to large negative values. Fig. 3a shows the source drain current $I_{SD}$ measured at $V_{SD}$ = 0.5 V, as a function of the potential of the reference electrode ($V_{ref}$), as is necessary to estimate the band gap (see below). The occurrence of hole conduction when $V_{ref}$ exceeds (in absolute value) -2 V is apparent.

We use the data of Fig. 3a to estimate the value of the band gap $\Delta$ of CVD-grown MoS$_2$ monolayers. To this end, we extract the values of the threshold voltages for electron $V_{th}^e$ and hole $V_{th}^h$ conduction from Fig. 3a, from which we obtain $\Delta = e(V_{th}^e - V_{th}^h)$ (see Supplementary Information and Ref. 39 for full details). Recent work on IL-gated transistors realized on WS$_2$ bulk[39], mono/bilayers[9], and bulk MoTe$_2$[40,41], has shown that this estimate is reliable to approximately 10% – or better – even when imperfections are present in the devices, mainly thanks to the extremely large geometrical capacitance of the IL gate. In MoTe$_2$ bulk devices, for instance, $\Delta = e(V_{th}^e - V_{th}^h)$ was found to agree with scanning tunneling data and optical spectroscopy data, even in the presence of significant unintentional doping and of sizable hysteresis in the device transfer curve[40]. Similarly to the case of bulk MoTe$_2$ devices, in our CVD-grown MoS$_2$ FETs the subthreshold swing for holes and electrons measured at $T$= 240 K is respectively 99 and 115 meV/dec (see Fig. 3b). These values, approximately twice larger than the ultimate limit $S = \frac{kT}{e}ln(10) = 47$ meV/dec, are indicative of contact non-ideality or of the presence of trap states inside the band gap. Indeed, in-gap states near the top of the valence band are visible in the left panel of Fig. 3b, where they manifest themselves in the enhancement of $I_{SD}$ measured deep in the hole subthreshold region (put in evidence by the dashed ellipse). Nevertheless, just as in the case of MoTe$_2$, it still makes sense to estimate the band gap from the device transfer curve. Depending on the specific device on which the analysis is done, we find band gap values $\Delta = e(V_{th}^e - V_{th}^h)$ ranging from 2.4 to 2.7 eV. When compared to existing estimates reported in the literature (2.15-2.5 eV, as obtained from scanning tunneling spectroscopy[42,43] and frequency-dependent photocurrent measurements[44]), our measurements contribute to constrain the true value of the band gap of MoS$_2$ monolayers. In this regard we emphasize that, when extracting the gap from the ambipolar FET transfer curves, it easy to understand how different physical processes lead to an overestimation of the value of $\Delta$, but not to an underestimation. Our experiments therefore strongly suggest that the actual value of $\Delta$ has to be 2.4 eV or smaller.

Having established the occurrence of ambipolar conduction in IL-gated CVD-grown MoS$_2$ transistors, we go back to the ambipolar injection regime and search for electroluminescence. Fig. 4a shows the output curve of one of our transistors measured at $V_G$ = 0.5 V, in which $V_{SD}$ is swept well past the onset of ambipolar injection (starting at $V_{SD}$ ~ 3 V). To detect light emission, we image the device under the microscope using a CCD camera. Images taken when the device is biased at increasingly larger $V_{SD}$ values are shown on the right side of Fig. 4a. For $V_{SD} < 4$ V, no emitted light is detected within the experimental sensitivity. For larger $V_{SD}$, electroluminescence from spots localized near hole-injecting contact becomes visible. If $V_{SD}$ is increased further, the light intensity emitted from these spots, as well as their number, increases. Light emission, however, continues to occur only near the interface with the hole-injecting contact.

The observed behavior is different from the one expected for ideal operation of light-emitting transistors[45], in which case the location where light is emitted from shifts in the channel upon varying $V_{SD}$. To ensure that emission does indeed occur from exciton recombination in the $MoS_2$ monolayer, and not from uncontrolled processes associated to carrier injection from the metal, it is therefore important to measure the electroluminescence energy spectrum. This is shown in Fig. 4b (red curve) together with the PL spectrum measured on the same devices, away from the contact region (blue curve; as discussed above – see Fig. 1d – the large sharp line at 690 nm and the smaller features just above 700 nm originate from PL of the sapphire substrate). The dominant peak in the electroluminescence spectrum coincides with the one in the PL spectrum, confirming that the light emitted from the bright spots indeed originates from exciton recombination in the $MoS_2$ monolayers. The electroluminescence spectrum shows an additional smaller and broader peak at approximately 770 nm (1.6 eV), whose origin is discussed in detail below (in the PL spectrum acquired at room temperature this peak is absent but it appears after cooling the sample to $T = 77$ K; see Supplementary Information Fig. S3).

At a first glance, the behavior of the observed electroluminescence appears similar to the phenomenology that has been reported previously[7] in devices realized using exfoliated monolayer MoS$_2$. A more careful inspection of the data, however, shows that important differences are present in the two cases, in particular in the evolution of EL spectrum upon increasing the applied bias. As we discuss below, the possibility to compare the gate-dependent ambipolar transport data with the bias dependent EL measurements in our devices provides strong evidence for a simple physical scenario that explains virtually all our observations. This scenario enables us to draw consistent conclusions about some aspects of the properties of CVD-grown MoS$_2$ monolayers, most notably of the in-gap states present at energies just above the top of the valence band.

To understand the behavior of the EL observed in our devices, it is important to realize that under the ambipolar injection conditions leading to light emission, transport occurs in the so-called space charge limited regime[46,47] for both electrons and holes. In this regime the charge carriers responsible for the current flow are not accumulated by the applied gate voltage, but are injected from the contacts (electrons at one contact and holes at the other), due to the large applied bias. As it is well established, the space charge limited transport regime is extremely sensitive to defects acting as trap for charge carriers, whose presence determines the profile of the electrostatic potential inside the material, and therefore the spatial distribution of the injected charge. In particular, a large density of "deep traps" – i.e., traps having an energy many times larger than $kT$ – can result in the spatial localization of carriers close to the injecting contact at low bias, and charge can penetrate deeper into the material only when a sufficiently high bias is applied (if the density of deep traps is large, the bias required to induce complete detrapping may be too high to be reached in the experiments without causing device failure) [46–48]. These considerations are relevant here because, as mentioned earlier, the measurement of $I_{SD}$ as a function of $V_{ref}$ (Fig. 3a) shows that a large density of states is present in the gap at an energy approximately 250-300 meV above the top of the valence band. At room temperature, these states act as traps for holes with a trapping energy of *10 x kT* or larger (i.e., they are deep traps), and cause the space charge due to the injected holes to remain localized close to the injecting contact. This explains why the position where electroluminescence originates from does not shift significantly upon increasing $V_{SD}$.

In the spectrum of the measured EL, the effect of trapping manifests itself in the shallow peak seen at energies below the peak at 660 nm originating from the recombination of "free" excitons, as usually observed in $MoS_2$ (see the red line in Fig. 4b). Indeed, the broad peak is 250-300 meV lower in energy than the free exciton peak, and this difference corresponds quite precisely to the energy of the trap states determined by the analysis of $I_{SD}$ –vs– $V_{ref}$ shown in Fig. 3a (in simple terms, the broad low-energy peak is due to excitons formed by an electron and a trapped hole that emits a photon of lower energy when recombining). The dependence of the amplitude of broad peak on $V_{SD}$ – shown in Fig. 4c – is consistent with the trapping scenario that we have outlined: as the applied bias increases, the amplitude of the shallow peak decreases relatively to the main "free" exciton peak because a larger $V_{SD}$ tends both to fill trap states and to de-trap holes, thereby increasing the number of free excitons relative to that of trapped ones. Interestingly, the trend shown in Fig. 4c leads to a sharpening of the spectrum upon increasing $V_{SD}$. This behavior is qualitatively the opposite of what has been observed by Sundaram et al.[7] in the EL of exfoliated $MoS_2$ monolayers, where the spectrum of the emitted light broadened very significantly in energy upon increasing $V_{SD}$. The large bias-induced broadening was attributed to hot photoluminescence, a conclusion that certainly appears plausible in those devices, because the power-per-unit-surface injected under the biasing conditions needed to observe EL was approximately $10^4$ times larger than in our experiments. These considerations illustrate how electroluminescence in MoS$_2$ monolayers can exhibit different behavior depending on the transport regime. It is only through the careful combined analysis of detailed aspects of the transport and opto-electronic processes – which in our devices is made possible by the use of an ionic liquid gate – that a consistent microscopic picture of the properties of MoS$_2$ monolayers can be established.

In conclusion, we have demonstrated the possibility to realize ionic-liquid gate field-effect transistors on large-area MoS$_2$ monolayers grown by chemical vapor deposition. Characterization of the material uniformity, of the transistor electrical characteristics, as well as the observation of electroluminescence show that the properties of devices realized on CVD-grown monolayers are as good as those of existing devices based on exfoliated material, and in several regards better. The observation of clear ambipolar conduction in the transfer curves of FETs, which had not been previously reported for exfoliated MoS$_2$ monolayers, allows us to extract specific information about the electronic properties of the material, including the size of the band gap and the identification of the in-gap states acting as trap for holes. Thanks to this information we have established a clear physical scenario that accounts in a consistent way for the results of transport and electroluminescence measurements. Finally, these results have technological relevance because the use of semiconducting monolayers transition metal dichalcogenides in future opto-electronic applications will necessarily rely on devices fabricated on large-area material produced by a controllable and scalable technique.

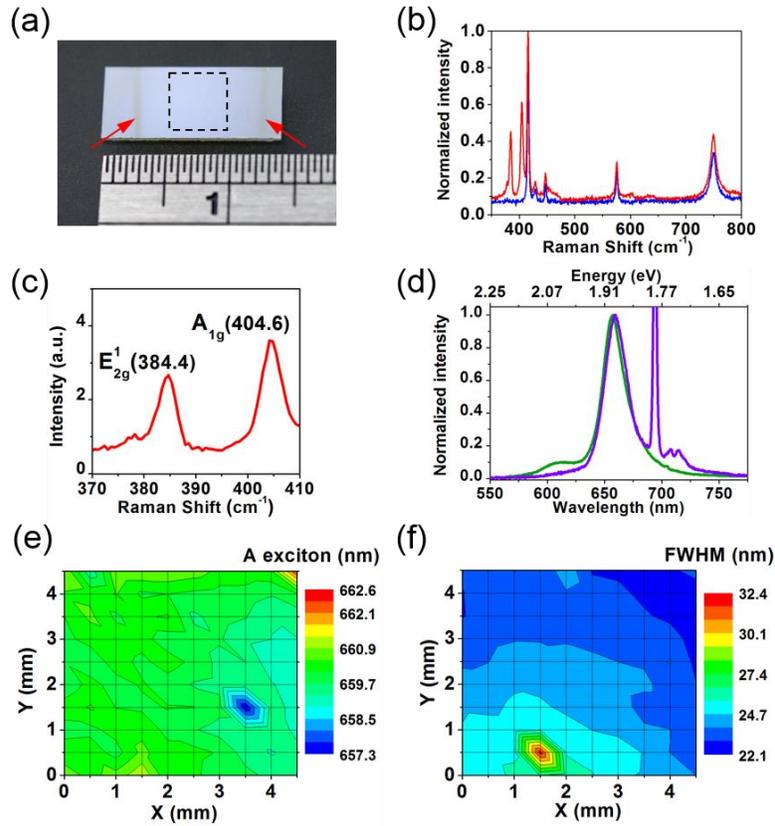

**Figure 1.** Raman and photoluminescence spectra of CVD-grown MoS$_2$ monolayers. (a) Optical image of CVD-grown MoS$_2$ on a sapphire substrate. The red arrows point to the edges of the MoS$_2$ monolayer; the dashed square delimitates the area on which the Raman and photoluminescence spectra shown in (e) and (f) were measured. (b) Raman spectra of a CVD-grown MoS$_2$ monolayer on sapphire (red) and of the bare sapphire substrate (blue). (c) Portion of the Raman signal originating from CVD-grown MoS$_2$, exhibiting the characteristic $E^1_{2g}$ and $A_{1g}$ peaks found in monolayers. (d) Comparison between the photoluminescence spectra of CVD-grown (violet) and exfoliated (green) monolayer MoS$_2$. The peaks observed in the CVD monolayer at about 1.78 eV and lower energies originate from the sapphire substrate (the peak at about 2.05 eV seen in exfoliated MoS$_2$ is due to B-exciton recombination). Maps of the A exciton position (e) and full width at half maximum (f) extracted from the photoluminescence spectra measured on CVD-grown MoS$_2$.

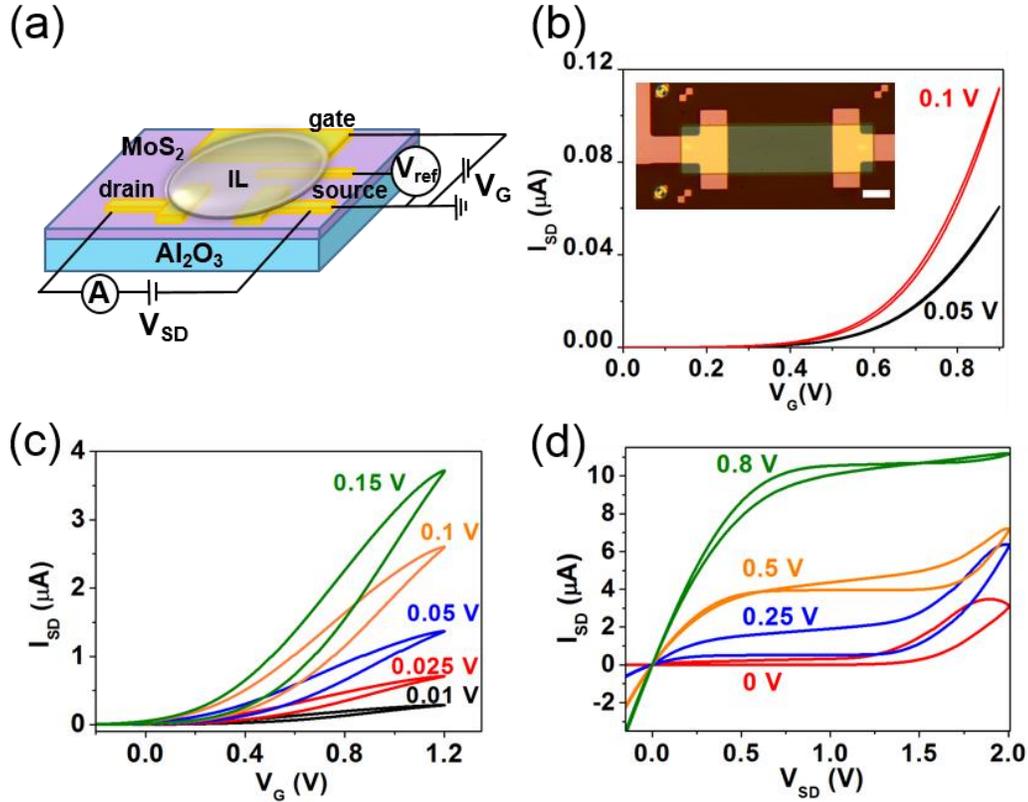

**Figure 2.** Electrical characteristics of monolayer MoS$_2$ ionic liquid-gated field-effect transistors. (a) Scheme of an ionic liquid-gated field-effect transistor (FET) realized using an as-synthesized MoS$_2$ monolayer, showing the source and drain contacts, the side gate and the reference electrode. (b) Source-drain current I$_{SD}$ under electron accumulation measured for two values of source-drain bias $V_{SD}$ as a function of gate voltage $V_G$ (FET transfer curves). The inset shows an optical microscope image of a device, in which the channel is defined by a window made in a PMMA layer (so that the IL is in contact with the MoS$_2$ only in this region). The scale bar in the inset corresponds to 20 μm. (c) Transfer curves of the same device shown in (b), showing a larger gate sweep range (that ultimately results in the occurrence of hysteresis, as commonly seen in devices based on exfoliated MoS$_2$ monolayers). (d) $I_{SD}$ as a function of $V_{SD}$ for different $V_G$ values (FET output curves) showing the typical behavior observed in an ambipolar transistor (i.e., an upturn in $I_{SD}$ is seen at sufficiently large $V_{SD}$, past the saturation regime).

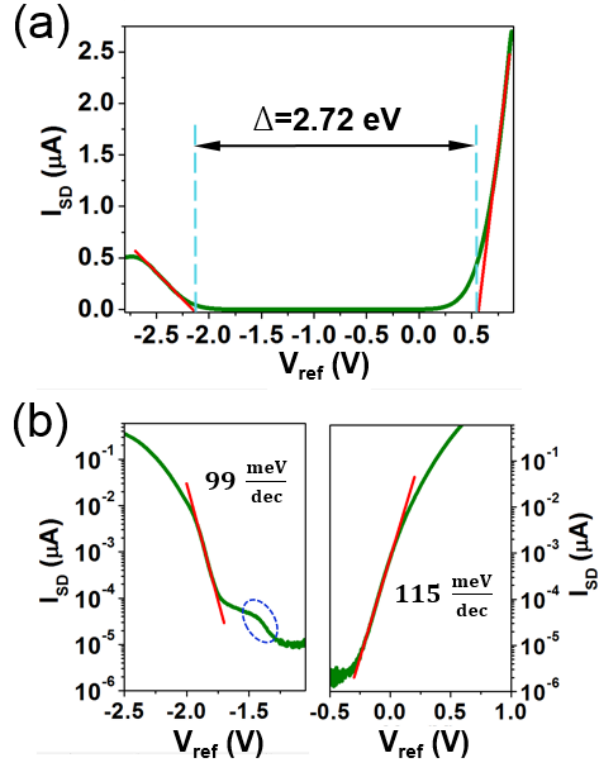

**Figure 3.** Estimation of the band gap of CVD grown monolayer MoS$_2$ from the transfer curves of an ionic-liquid gated transistor. (a) Source-drain current $I_{SD}$ as a function of reference voltage $V_{ref}$, measured for $V_{SD}$ = 0.5 V. The red lines are an extrapolation of the linear regime to $I_{SD}$ = 0 V, as needed to extract the threshold voltage (whose position for holes and electrons is indicated by the vertical blue-dashed lines). The difference between the threshold voltage of holes and electrons provides an estimate of the band gap ($\Delta$ = 2.7 eV in the present case). (b) Semi logarithmic plots of the data shown in (a). The red lines are the linear regressions made to estimate the sub-threshold swing. The dashed ellipse emphasizes the presence of an anomalously large contribution to the source drain current very deep in the hole subthreshold regime, due to in-gap states at energies close to the top of the valence band.

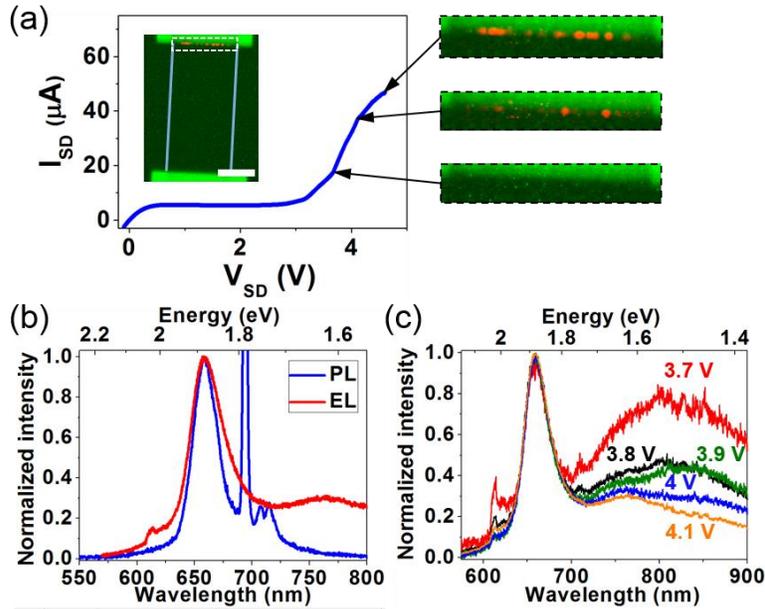

**Figure 4.** Electroluminescence in a CVD-grown MoS$_2$ monolayer ionic-liquid FET. (a) Source-drain current $I_{SD}$ as a function of source-drain bias extending to large $V_{SD}$ values, as it is needed to drive the device into the light-emission regime. The inset shows an image of the device with the region of the channel close to the hole injecting contact – where light emission occurs – delimited by a white dashed line. The three panels on the right show images of this region for three different values of bias (pointed to by the black arrows). Light emitting spots (red spots in the image) appear upon increasing $V_{SD}$. The scale bar in the inset corresponds to 20 μm. (b) Comparison between the photoluminescence (blue curve) and electroluminescence (red curve) spectra ($V_{SD}$ =4.1 V) measured in the same device, showing that the electroluminescence peaks at the energy expected from the recombination of A excitons in MoS$_2$ (the intensities of both curves are normalized to the A-exciton peak). A smaller and broader peak is seen in the EL spectrum originating from trapped excitons; in PL the same peak is observed but only at lower temperature (see Supplementary Information). (c) Normalized electroluminescence spectra recorded for different values of source-drain bias $V_{SD}$ demonstrating the decrease of the lower energy peak with upon increasing bias, consistently with what is expected for trapped excitons.

## ASSOCIATED CONTENT

**Supporting Information**

Hall effect measurements, description of the band gap extraction from FET transfer curves, and low-temperature photoluminescence measurements. This material is available free of charge via the Internet at http://pubs.acs.org

## AUTHOR INFORMATION

**Corresponding Author**

*E-mail: Alberto.Morpurgo@unige.ch.

**Author Contributions**

EP grew the $MoS_2$ monolayers and performed the transport and optical measurements with the help of IGL (transport experiments) and NU (optical characterization). AFM planned and directed the research. All authors discussed the data, contributed to their interpretation, and participated in writing the manuscript. All authors have given approval to the final version of the manuscript.

**Funding Sources**

We gratefully acknowledge the Swiss National Science Foundation (SNF) and the EU Graphene Flagship Project for financial support.

**Notes**

The authors declare no competing financial interest.

ACKNOWLEDGMENT

We are very grateful to A. Ferreira for his constant technical support throughout this work.